\documentclass[prl,twocolumn,superscriptaddress,showpacs]{revtex4}

\usepackage{xcolor}
\usepackage{soul}
\usepackage{graphicx}
\usepackage{dcolumn}
\usepackage{bm}
\usepackage{color}

\def\up{\uparrow}
\def\dwn{\downarrow}

\def\gesim{\ \raise.3ex\hbox{$>$}\kern-0.8em\lower.7ex\hbox{$\sim$}\ }

\begin{document}
\title{Mechanism of screening or enhancing the pseudogap
throughout the two-band Bardeen-Cooper-Schrieffer to Bose-Einstein condensate crossover
}
\author{Hiroyuki Tajima}
\affiliation{Department of Mathematics and Physics, Kochi University, Kochi 780-8520, Japan}
\affiliation{RIKEN Nishina Center, Wako, Saitama, 351-0198, Japan}
\author{Yuriy Yerin}
\affiliation{Dipartimento di Fisica e Geologia, Universit\`{a} di Perugia, I-06123 Perugia, Italy}
\affiliation{School of Science and Technology, Physics Division, Universit\`{a} di Camerino, 62032 Camerino (MC), Italy}
\author{Pierbiagio Pieri}
\affiliation{Dipartimento di Fisica e Astronomia, Universit\`{a}  di Bologna, I-40127 Bologna, Italy}
\affiliation{INFN, Sezione di Bologna, I-40127 Bologna, Italy}
\author{Andrea Perali}
\affiliation{School of Pharmacy, Physics Unit, Universit\`{a} di Camerino, 62032 Camerino (MC), Italy}
\date{\today}
\begin{abstract}
We demonstrate the rise-and-fall of multiple pseudogaps in the Bardeen-Cooper-Schrieffer-Bose-Einstein-condensation (BCS-BEC) crossover in two-band fermionic systems having different pairing strengths in the deep band and in the shallow band.
The striking features of this phenomenon are an unusual many-body screening of pseudogap state and the importance of pair-exchange couplings, which induces multiple pseudogap formation in the two bands.
The multi-band configuration suppresses pairing fluctuations and the pseudogap opening in the strongly-interacting shallow band at small pair-exchange couplings by screening effects, with possible connection to the pseudogap phenomenology in iron based superconductors.
On the other hand, the multiple pseudogap mechanism accompanies with the emergence of binary preformed Cooper pairs originating from interplay between intra-band and pair-exchange couplings.
\end{abstract}
\maketitle
The discovery of unconventional superconductors, which started with heavy fermions, followed by organic superconductors, and then by cuprate compounds, has prompted an era of tremendous growth of activities in condensed matter research~\cite{Scalapino,Zehetmayer}. The complex structure of the order parameter in these systems brought a plethora of unique phenomena and effects, with no counterparts in conventional superconductors, such as a broken time-reversal symmetry, collective modes, and an unusual Josephson effect~\cite{Lin2014,Tanaka2015,Yerin2017}. 
The new degrees of freedom in multi-component and multi-band superconductors has been anticipated to be a promising root toward the realization of room-temperature superconductivity~\cite{Milosevic}.
Such unconventional superconductors can exhibit anomalous normal state characteristics above their critical temperature $T_{\rm c}$, which are interpreted as the pseudogap state~\cite{Fischer,Mueller}, corresponding to the presence of gap-like features above $T_{\rm c}$ but with a finite spectral intensity at low frequencies~\cite{note1}. 
The origin of the pseudogap is a key for understanding of the pairing glue in unconventional superconductors.
Pseudogap effects have also been discussed in the context of the Bardeen-Cooper-Schrieffer (BCS) to Bose-Einstein condensation (BEC) crossover,
where the BCS state of overlapping Cooper pairs changes continuously to the BEC of tightly bound molecules with increasing attractive interaction~\cite{Eagles,Leggett,Nozieres,SadeMelo,Perali2,Ohashi,Giorgini,Bloch,Strinati,OhashiTajimaWyk}. It is experimentally achieved in ultracold Fermi atomic gases exploiting  Fano-Feshbach resonances~\cite{Regal,Bartenstein,Zwierlein}. Also ultracold Fermi gases in the BCS-BEC crossover regime exhibit strong pairing fluctuations and pseudogap effects~\cite{Tsuchiya,NP2010,PRL11,Palestini2012}.
\par
Among unconventional superconductors, the recently discovered iron-based superconducting compounds attract attention, since some of them are expected to place in the BCS-BEC crossover regime due to their large ratio between the superconducting gap and the Fermi energy~\cite{Okazaki,Kasahara,Kasahara2,Rinott}.
This new class of superconductors opens a new frontier for the study of the multi-band BCS-BEC crossover, where non-trivial features have been discussed~\cite{Perali1996,Bianconi1998,Iskin4,Iskin,He2009,Guidini,Takahashi2014,He2015,Chubukov,Wolf,Salasnich,Klimin,Klimin2,Vargas}. 
Like for other unconventional superconductors, there is now expanding experimental evidence that the pseudogap is realized in iron-based compounds~\cite{Kwon2012,Matusiak2015,Seo2019,Kang2020,Solovjov},
despite some reports about the missing of strong pairing fluctuations and pseudogap effects~\cite{Hanaguri,Takahashi}. 
In order to understand the controversial pseudogap physics in multiband and multicomponent systems like iron-based superconductors, a unified description of the multi-band BCS-BEC crossover is required. Such a theory can be useful to describe also many-body physics in Yb Fermi gases near the orbital Feshbach resonance~\cite{Zhang,Pagano,Hofer,Iskin3,He2016,Xu,Iskin2,Zou,Mondal}, thus bridging these atomic systems with multiband superconductors. The multi-channel many-body theory is also of importance to unveil pairing properties in nanostructured superconductors~\cite{Perali1996,Bianconi1998,Chen} and electron-hole systems~\cite{Tomio,Pieri,Perali2013}.
\par
In this article, we develop a theory of the two-band BCS-BEC crossover in the normal state above $T_{\rm c}$ based on the $T$-matrix approach~\cite{TajimaCM}, which has been successfully applied to strongly interacting attractive Fermi gases~\cite{Pini}.
We address the single-particle density of states (DOS) and elucidate competing mechanisms of screening and enhancement of the pseudogap  in two-band systems.
The screening of pairing fluctuations and resulting reduction of the pseudogap regime are found in our results at the unitarity limit of the shallow band for weak pair-exchange couplings.
{This result suggests that, in the two-band system, the Fulde-Ferrel-Larkin-Ovchinnikov state~\cite{FF,LO}, tending to be disrupted by pairing fluctuations~\cite{Shimahara,OhashiFF}, 
is more stable compared to the single-band case, as observed in recent experiments~\cite{Kasahara2020,Molatta}.}

On the other hand, the strong pair-exchange coupling leads to multiple pseudogap and the emergence of binary preformed Cooper pairs in the crossover regime. 
This is in contrast with the pseudogap in ultracold Fermi gases, which is induced by strong intra-band couplings.
Hereafter, we take $\hbar=k_{\rm B}=1$ and unit volume.
\par
\begin{figure}[t]
\begin{center}
\includegraphics[width=7.5cm]{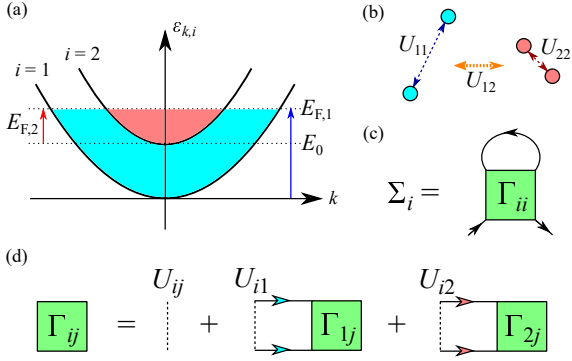}
\end{center}
\caption{(a) Two-band electronic structure considered in this work.
The two bands ($i=1,2$) are separated in energy by $E_0$.
Resulting Fermi energies $E_{{\rm F},i}$ have the relation $E_{\rm F,1}=E_{\rm F,2}+E_0$.
(b) Illustration of how the interactions $U_{ij}$ work in our configuration.
While $U_{11}$ and $U_{22}$ cause intra-band Cooper pairing in each band, $U_{12}$ ($=U_{21}$) introduces pair-tunneling between the two bands.
(c) and (d) show Feynman diagrams for the self-energy $\Sigma_i$ and the multi-band $T$-matrix $\Gamma_{ij}$ in our $T$-matrix approach, respectively.
}
\label{fig1}
\end{figure}
As shown in Fig.~\ref{fig1}(a), we consider a two-band model where the second shallow band ($i=2$) is coupled with the first deep band ($i=1$)~\cite{Tajima,Yerin}, as described by the Hamiltonian~\cite{SMW}
\begin{equation}
\label{eq1}
H=\sum_{\bm{k},\sigma,i}\xi_{\bm{k},i}c_{\bm{k},\sigma,i}^{\dag}c_{\bm{k},\sigma,i} +\sum_{i,j}U_{ij}\sum_{\bm{q}}B_{\bm{q},i}^{\dag}B_{\bm{q},j},
\end{equation}
where $\xi_{\bm{k},i}=k^2/(2m_i)-\mu+E_0\delta_{i,2}$ is the kinetic energy measured from the chemical potential $\mu$ with the energy separation $E_0$ between two bands and $\delta_{i,2}$ is the Kronecker delta.
We use equal effective masses $m=m_1=m_2$, for simplicity.
$c_{\bm{k},\sigma,i}$ and $B_{\bm{q},i}=\sum_{\bm{k}}c_{-\bm{k}+\bm{q}/2,\dwn,i}c_{\bm{k}+\bm{q}/2,\up,i}$ are spin-$\sigma=\up,\dwn$ fermion and spin-singlet pair annihilation operators in the $i$-band, respectively.
In this work, we use $E_0=0.6E_{\rm F,1}$ where $E_{{\rm F},i}=(3\pi^2n_i)^{\frac{2}{3}}/(2m)$ is the non-interacting Fermi energy in the $i$-band, defined in terms of the number density $n_i$.
The intra-band couplings $U_{ii}$ can be characterized in terms of the intra-band scattering lengths $a_{ii}$ as
\begin{eqnarray}
\frac{m}{4\pi a_{ii}}=\frac{1}{U_{ii}}+\sum_{\bm{k}}^{k_0}\frac{m}{k^2},
\end{eqnarray}
where $k_0$ is the momentum-cutoff taken to be $100k_{\rm F,t}$. Here, $k_{\rm F,t}\equiv\sqrt{2m E_{\rm F,t}}$ is the Fermi wavevector associated with the total Fermi energy $E_{\rm F,t}=(3\pi^2 n)^{2/3}/(2m)$, defined in terms of the total number density $n$. In a similar way, one defines the Fermi wavevectors $k_{{\rm F}, i}$ in each band, which are used to define the dimensionless intra-band coupling strengths  $(k_{\rm F,1}a_{11})^{-1}$ and $(k_{\rm F,2}a_{22})^{-1}$.
In this work, we use $(k_{\rm F,1}a_{11})^{-1}\leq -2$ and $-1\leq (k_{\rm F,2}a_{22})^{-1}\leq1$.
With this choice of couplings, pairs forming in the deep band ($i=1$) have a BCS character, while the BCS-BEC crossover is tuned in the shallow band ($i=2$).
{Although we consider the 3D system, it is expected to be relevant to FeSe multi-band superconductors since recent experiments exhibit a 3D wave-vector dependence of the superconducting gap~\cite{Kushnirenko}, indicating that a 3D theoretical approach is applicable.
In addition, the strong-coupling regime from the unitarity to the BEC side in 3D would be similar to the 2D counterpart due to the presence of the two-body bound state. }
For convenience, we also introduce a dimensionless pair-exchange coupling 
$\lambda_{12}=U_{12}(k_0/k_{\rm F,t})^2n/E_{\rm F,t}$ where $U_{21}=U_{12}$~\cite{Tajima,Yerin}. 
\par
The $i$-band self-energy in the multi-band $T$-matrix approach reads
\begin{equation}
\label{eqsig}
\Sigma_i(\bm{k},i\omega_s)=T\sum_{\bm{q},i\nu_l} \Gamma_{ii}(\bm{q},i\nu_l)G^{0}_{i}(\bm{q}-\bm{k},i\nu_l-i\omega_s),
\end{equation}
where $\omega_s=(2s+1)\pi T$ and $\nu_l=2l\pi T$ ($s$ and $l$ integer) are fermionic and bosonic Matsubara frequencies, respectively.
$G_i^0(\bm{k},i\omega_s)=[i\omega_s-\xi_{\bm{k},i}]^{-1}$ is the bare Green's function.
The many-body $T$-matrix $\{\Gamma_{ij}\}_{2\times 2}$, which sums up the ladder-type diagram shown in Fig.~\ref{fig1}(d), is given by
\begin{eqnarray}
\Gamma_{ij}(\bm{q},i\nu_l)=U_{ij}+\sum_{\ell=1,2}U_{i\ell}\Pi_{\ell\ell}(\bm{q},i\nu_l)\Gamma_{\ell j}(\bm{q},i\nu_l),
\end{eqnarray}
where $\Pi_{\ell\ell}$ is 
\begin{eqnarray}
\Pi_{\ell \ell}(\bm{q},i\nu_l)=-T\sum_{\bm{p},i\omega_s}G_i^0(\bm{p}+\bm{q},i\omega_s+i\nu_l)G_i^0(\bm{p},-i\omega_s).
\end{eqnarray}
Fixing $\mu$ by solving the number equation $n=n_1+n_2$ with
\begin{equation}
\label{eq9}
n_{i}=2T\sum_{\bm{k},{i\omega_s}}G_{i}(\bm{k},i\omega_s),    
\end{equation}
where $G_i(\bm{k},i\omega_s)=[i\omega_s-\xi_{\bm{k},i}-\Sigma_i(\bm{k},i\omega_s)]^{-1}$ is the dressed Green's function,
we obtain the superfluid/superconducting critical temperature $T_{\rm c}$ from the Thouless criterion~\cite{Thouless}
$\left[\Gamma_{22}(\bm{q}=0,i\nu_l=0)\right]^{-1}=0$.
[While, in the presence of $U_{12}$, all the matrix elements $\Gamma_{ij}(\bm{q}=0,i\nu_l=0)$ diverge simultaneously at $T_{\rm c}$,
in the case of vanishing $U_{12}$  only $\Gamma_{22}$ diverges, due to our choice of the coupling strengths.]
{We numerically evaluated the Matsubara frequency sum in Eqs. (\ref{eqsig}) and (\ref{eq9}) with finite cutoffs~\cite{Pieri2004} and checked their convergences~\cite{Supplement}.}
\par
The DOS is obtained from 
\begin{eqnarray}
N_{i}(\omega)=-\frac{1}{\pi}\sum_{\bm{k}}{\rm Im}G_i(\bm{k},i\omega_s\rightarrow \omega+i\delta),
\end{eqnarray}
where we take $\delta=O(10^{-3})E_{\rm F,t}$.
{For simplicity}, the analytic continuation is numerically performed by using the method of Pad\`{e} approximants~\cite{Supplement,Serene} {(see Supplemental Material).}
\par
\begin{figure}[t]
\begin{center}
\includegraphics[width=7.5cm]{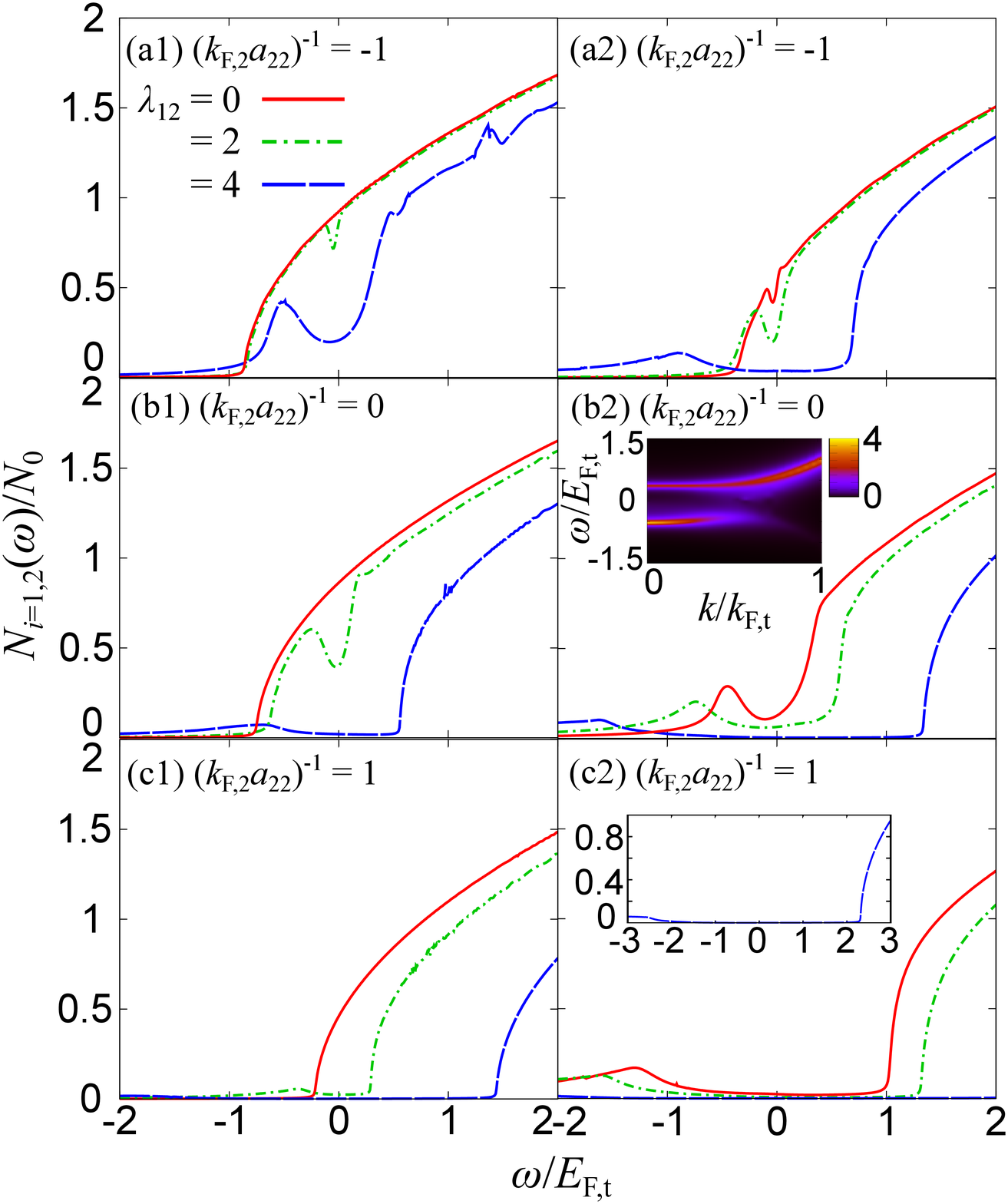}
\end{center}
\caption{DOS $N_{i=1,2}(\omega)$ in the multi-band BCS-BEC crossover.
The left (right) panels show $N_1(\omega)$ ($N_2(\omega)$) at weak coupling $(k_{\rm F,2}a_{22})^{-1}=-1$ [(a1), (a2)], unitarity $(k_{\rm F,2}a_{22})^{-1}=0$ [(b1), (b2)], and strong coupling $(k_{\rm F,2}a_{22})^{-1}=1$ [(c1), (c2)]. In all panels, we fix $(k_{\rm F,1}a_{11})^{-1}=-4$.
The dimensionless pair-exchange coupling is taken as $\lambda_{12}=0$, $2$, and $4$.
For reference, we present the spectral weight $A_{2}(\bm{k},\omega)E_{\rm F,t}=-{\rm Im}G_{2}(\bm{k},\omega+i\delta)E_{\rm F,t}/\pi$ at $\lambda_{12}=0.5$ in the inset of panel (b2).
The inset of (c2) shows $N_2(\omega)$ at $\lambda_{12}=4$ because of the large energy gap.
$N_0=mk_{\rm F,t}^2/(2\pi^2)$ is the non-interacting DOS associated with total number density $n$. 
}
\label{fig2}
\end{figure}
Figure~\ref{fig2} shows the DOS $N_i(\omega)$ in the multi-band BCS-BEC crossover.
In the case of $\lambda_{12}=0$, 
{while the small intraband coupling in the deep band $(k_{\rm F,1}a_{11})^{-1}=-4$ does not suppress a square-root behavior typical of non-interacting gases, $N_0(\omega)\propto \sqrt{\omega+\mu}$,} 
$N_2(\omega)$ exhibits the pseudogap around $\omega=0$~\cite{Trivedi1995} due to strong pairing fluctuations associated with $U_{22}$.
It is consistent with the results obtained in the single-band counterpart.
On the other hand, in the presence of the non-zero pair-exchange coupling, $N_1(\omega)$ also shows the pseudogapped DOS even with the weak intraband coupling.
This is thus a ``pair-exchange-induced pseudogap".
In addition, the coupling $\lambda_{12}$ enlarges the pseudogap in $N_2(\omega)$.
{The} pair-exchange-induced pseudogap in $N_{1}(\omega)$ becomes larger when the intra-band coupling in the shallow band $(k_{\rm F,2}a_{22})^{-1}$ gets stronger.
Eventually, at very strong pair-exchange coupling such as $\lambda_{12}=4$, both $N_{1}(\omega)$ and $N_{2}(\omega)$ show a fully-gapped structure due to the large two-body binding energy.

These features can be qualitatively understood as follows. Quite generally, the size of pseudogap effects in the  band $i$ can be roughly estimated by the energy scale  $\Delta_{\infty,i}^2=-T\sum_{\bm{q},i\nu_l}\Gamma_{ii}(\bm{q},i\nu_l)$ introduced in Ref.~\cite{PieriNP} for a single band, and here generalized to the multiband case. Even though in general $\Delta_{\rm \infty}$  is related to the so-called Tan's contact $C$~\cite{Tan}, it was shown in Ref. \cite{PalestiniC} that in the intermediate crossover regime and close to $T_{\rm c}$, $\Delta_{\rm \infty}$ is close to the pseudogap scale energy determined from $N(\omega)$.
In the two-band case in the presence of a finite $\lambda_{12}$,  $\Gamma_{11}(\bm{q},i\nu_l)$ and $\Gamma_{22}(\bm{q},i\nu_l)$  diverge simultaneously at $T_{\rm c}$, for  $\bm{q}=0$ and $\nu_l=0$.
For this reason, the scales $\Delta_{{\rm \infty},1}$ and $\Delta_{{\rm \infty},2}$ become interconnected, explaining in this way the pair-exchange-induced pseudogap in the deep band.
\par
\begin{figure}[t]
\begin{center}
\includegraphics[width=6cm]{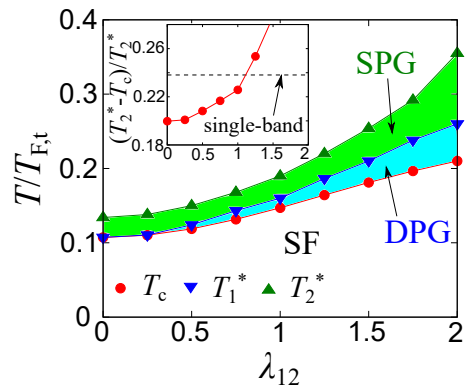}
\end{center}
\caption{The band-dependent pseudogap temperatures $T_{i=1,2}^*$ and the critical temperature $T_{\rm c}$ as functions of $\lambda_{12}$.
The intra-band couplings are chosen as $(k_{\rm F,2}a_{22})^{-1}=0$ and $(k_{\rm F,1}a_{11})^{-1}=-2$.
The regions where $T_{\rm c}<T<T_{\rm 1}^*$ and $T_{\rm 1}^*<T<T_{\rm 2}^*$ are double-pseudogap (DPG) and single-pseudogap (SPG) regimes, respectively. 
The inset shows the ratio $(T_{2}^*-T_{\rm c})/T_{\rm c}$ as a function of $\lambda_{12}$ which characterizes how the pseudogap regime in the shallow band is shrunk by multi-band effects.
The horizontal dashed line in the inset shows the single-band counterpart.
}
\label{fig3}
\end{figure}
To characterize the pseudogap state, we introduce the band-dependent pseudogap temperatures $T_{i=1,2}^*$ where the minimum of $N_i(\omega)$ around $\omega=0$ disappears~\cite{Tsuchiya}.
Figure~\ref{fig3} shows the obtained phase diagram at unitarity (crossover regime) of the shallow band coupled with the weakly interacting deep band, where $(k_{\rm F,2}a_{22})^{-1}=0$ and $(k_{\rm F,1}a_{11})^{-1}=-2$.
In this figure, we plot the critical temperature $T_{\rm c}$ and pseudogap temperatures $T_{1,2}^*$ as functions of $\lambda_{12}$.
While the single pseudogap (SPG) appears in the region $T_1^*<T<T_2^*$,
the double pseudogaps (DPG) can be found below $T=T_1^*$.
In the case of vanishing $\lambda_{12}$, since the deep band does not exhibit pseudogap behavior, we obtain $T_{1}^*=T_{\rm c}$.
However, if $\lambda_{12}$ is shifted from zero to strong coupling,
$T_{1}^*$ deviates from $T_{\rm c}$ due to the interband pairing fluctuations.
Thus, the pseudogap regime in the deep band ($T_{\rm c}<T<T_2^{*}$) originates purely from the pseudogap induced by the transfer of pair-fluctuations due to the pair-exchange (rise of induced pseudogap). 
\par
The inset of Fig.~\ref{fig3} shows the ratio $(T_2^*-T_{\rm c})/T_{\rm c}$ as a function of $\lambda_{12}$.
For a reference, we plot in this figure the numerical value obtained in the single-band counterpart at the unitarity limit. 
The pseudogap regime ($T_{\rm c}<T<T_{2}^*$) in the two-band case with small $\lambda_{12}$ is clearly reduced compared to the single-band counterpart (fall of pseudogap). 
This tendency is consistent with the experiments for FeSe multi-band superconductors in the BCS-BEC crossover regime~\cite{Hanaguri,Takahashi} as well as with previous theoretical work~\cite{Salasnich,Tajima}.
This screening effect is related to the Pauli-blocking produced by the large Fermi surface in the deep band for our two-band configuration~\cite{Guidini}.
However, such a regime is destroyed if one shifts $\lambda_{12}$ to the strong-coupling regime ($\lambda_{12}\gesim 1$) due to strong interband pairing fluctuations.
\par
\begin{figure}[t]
\begin{center}
\includegraphics[width=6cm]{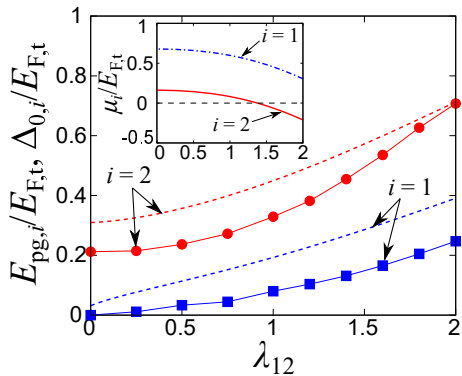}
\end{center}
\caption{The pseudogap sizes $E_{{\rm pg},i}$ estimated from the single-particle DOS at $T=T_{\rm c}$  (symbols) are compared with the mean-field gaps $\Delta_{0,i}$ at $T=0$ (dashed lines) as a function of the dimensionless pair-exchange coupling $\lambda_{12}$. The intraband interaction parameters are chosen as $(k_{\rm F,2}a_{22})^{-1}=0$ and $(k_{\rm F,1}a_{11})^{-1}=-2$.
The inset shows the chemical potential $\mu_i \equiv\mu-E_0\delta_{i,2}$ referred to the bottom of each band.}
\label{fig4}
\end{figure}
Figure~\ref{fig4} shows a comparison between the pseudogap energies $E_{{\rm pg},i}$ obtained from our $T$-matrix approach at $T=T_{\rm c}$ and the mean-field gaps $\Delta_{0,i}$ at $T=0$~\cite{Yerin}.
Here, $E_{\rm pg,i}$ is the half width of the dip structure in $N_i(\omega)$ around $\omega=0$. Specifically,  we define
$E_{{\rm pg},i}=(\omega_i'-\omega_{{\rm LM},i})/2$ where $\omega_{{\rm LM}, i}<0$ is the frequency where $N_i(\omega)$ has a local maximum due to the pseudogap and $\omega_i'>0$ is determined such that $N_i(\omega_i')=N_i(\omega_{{\rm LM},i})$~\cite{Tsuchiya,Watanabe}.
The dependence of $E_{{\rm pg},i}$ and  $\Delta_{0,i}$  on $\lambda_{12}$ are qualitatively  similar.
As for the single-band BCS-BEC crossover, the pseudogap can be regarded as half the energy needed to excite a single-particle by breaking a preformed Cooper pair. 
The coexistence and different magnitudes of the pseudogap energies $E_{\rm pg,1}$ and $E_{\rm pg,2}$ indicates the emergence of binary preformed Cooper pairs.
It is consistent with our prediction of binary molecular BEC with different pair sizes in the strong-coupling regime~\cite{Tajima}.
Indeed, different intraband pair-correlation lengths, corresponding to different Cooper pair size in each band, are obtained also within the mean-field approach at $T=0$~\cite{Yerin}.
In addition, this picture is supported by the emergence of binary Tan's contacts characterizing two kinds of pair correlations in the two-band system~\cite{TajimaCM}.
The finding that $E_{{\rm pg},i}$ is smaller compared to $\Delta_{0,i}$ is also consistent with the single-band result~\cite{Watanabe}.
We note that in the strong pair-exchange coupling regime $\lambda_{12}\gesim 1.5$, $\mu_2=\mu-E_0$ changes its sign {(where $\mu_{i}=\mu-E_0\delta_{i,2}$ is the chemical potential measured from the bottom of each band)} due to the large two-body binding energy associated with $U_{22}$ as well as with $\lambda_{12}$ (see the inset of Fig.~\ref{fig4}).
In such a regime, $N_i(\omega)$ exhibits a fully-gapped structure and $E_{{\rm pg},i}$ progressively approaches the two-body binding energy.
Although not shown here, $\mu_1$ also changes sign in the stronger coupling regime. 
\par
\begin{figure}[t]
\begin{center}
\includegraphics[width=6cm]{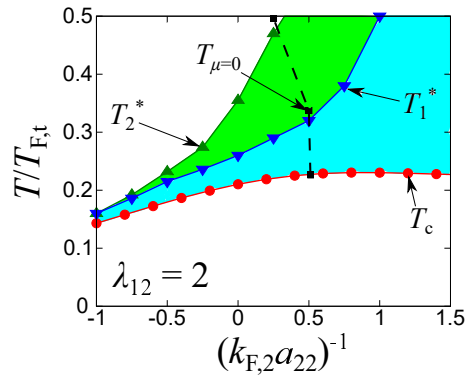}
\end{center}
\caption{Two-band BCS-BEC crossover phase diagram in the temperature vs intraband coupling $(k_{\rm F,2}a_{22})^{-1}$ plane for a  strong pair-exchange coupling $\lambda_{12}=2$ and $(k_{\rm F,1}a_{11})^{-1}=-2$. 
$T_{\mu=0}$ shows the temperature where $\mu=0$.}
\label{fig5}
\end{figure}
Finally, we report the phase diagram of the two-band BCS-BEC crossover for strong pair-exchange coupling $\lambda_{12}=2$, as shown in Fig.~\ref{fig5}.
At weak intra-band couplings, two pseudogaps simultaneously open in the two bands.  
These multiple pseudogaps originate from the strong pair-exchange coupling.
On the other hand, when the intraband coupling in the shallow band increases, the two pseudogap temperatures deviate from each other, indicating multiple energy scales of pseudogaps as shown in Fig.~\ref{fig4}. This multiple pseudogap regime evolves eventually into a molecular binary Bose gas regime.  Although the boundaries between these regimes are not sharp, the temperature $T_{\mu=0}$ at which the chemical potential $\mu$ goes below the bottom of the deep band could be used as a qualitative crossover line separating the two regimes at low temperature.
\par
In conclusion, we have demonstrated how multiple pseudogaps appear and when pair fluctuations are screened in the two-band BCS-BEC crossover at arbitrary pair-exchange couplings.
While the pair fluctuations inducing the pseudogap are screened by multi-band effects at weak pair-exchange couplings,
this screening regime turns into multiple pseudogaps at strong pair-exchange due to interband pairing fluctuations.
We have constructed the phase diagram of the two-pseudogap state in the temperature and pair-exchange plane, and show the pseudogap temperatures where single and multiple pseudogaps appear in the single-particle density of states.
Examining the pseudogap temperature in the shallow band, we have confirmed that the screening of pairing fluctuations due to the multi-band nature can be found in the BCS-BEC crossover regime.
Furthermore, the different magnitudes of the pseudogaps indicates the presence of binary preformed Cooper pairs with different binding energies and sizes, as also confirmed from the comparison between the pseudogap size at the critical temperature and the mean-field energy gaps at $T=0$. 
\par
We believe our results to be quite general: by relaxing, if required, some restrictions of the model considered here, such as the fixed energy shift and the electron-like character of bands,  the idea of multi-channel pairing fluctuations could be applied to a variety of strongly correlated multi-component systems such as cold atoms, electron-hole systems, nuclear matter, and nanostructured materials.
\par
{\it Acknowledgment}---
H. T. is grateful for the hospitality of the Physics Division at University of Camerino. 
H. T. was supported by Grant-in-Aid for JSPS fellows (No.17J03975) and for Scientific Research from JSPS (No.18H05406).

\pagebreak
\widetext
\begin{center}
\textbf{\large Supplemental Materials: Mechanism of screening or enhancing the pseudogap
throughout the two-band Bardeen-Cooper-Schrieffer to Bose-Einstein condensate crossover}
\end{center}
\setcounter{equation}{0}
\setcounter{figure}{0}
\setcounter{table}{0}
\setcounter{page}{1}
\makeatletter
\renewcommand{\theequation}{S\arabic{equation}}
\renewcommand{\thefigure}{S\arabic{figure}}
\renewcommand{\bibnumfmt}[1]{[S#1]}
\renewcommand{\citenumfont}[1]{S#1}

\section{Analytic continuation with the Pad\'{e} approximants}
\label{secs1}
\begin{figure}[h]
\begin{center}
\includegraphics[width=7cm]{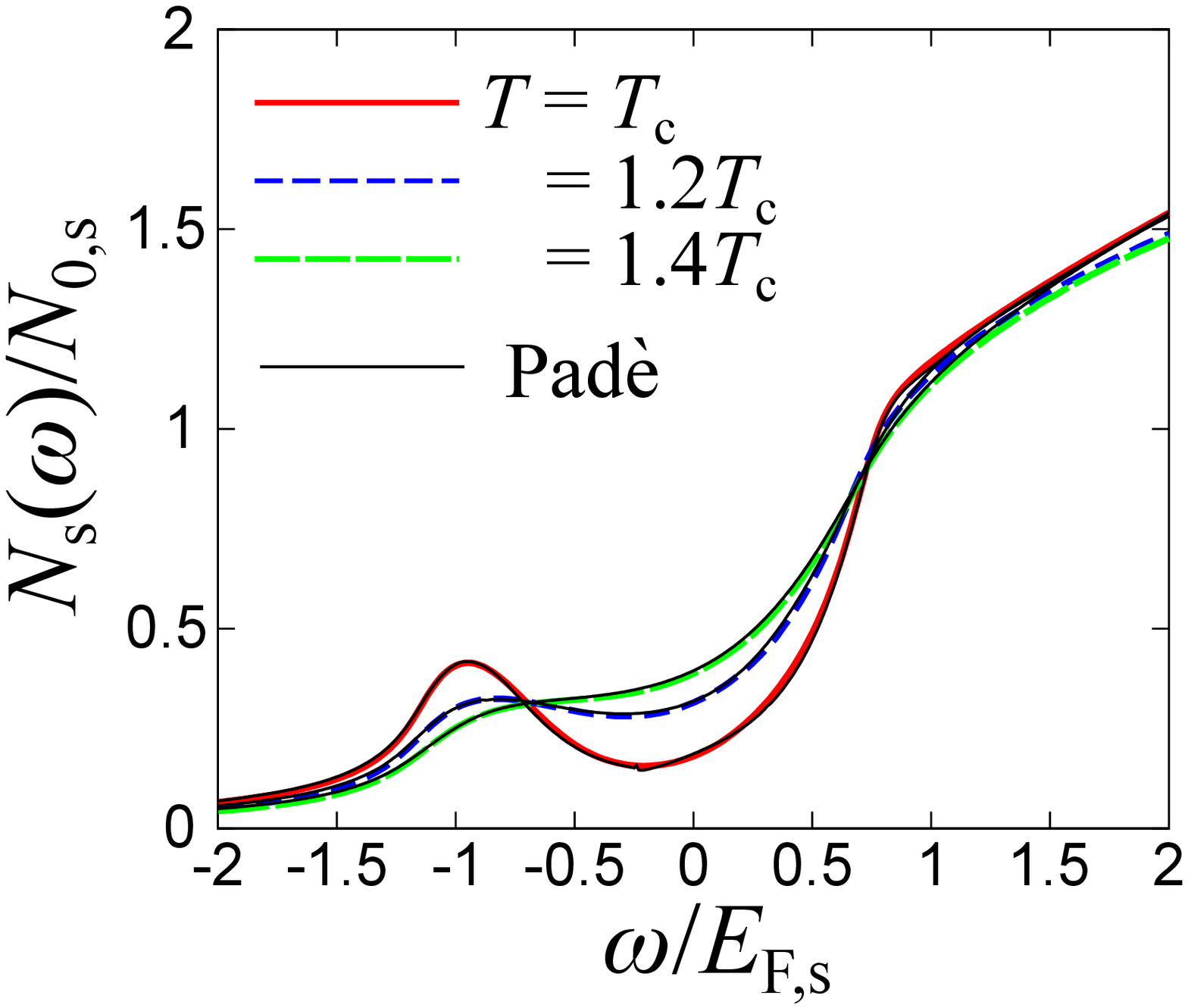}
\end{center}
\caption{Comparison of the DOS $N_{\rm s}(\omega)$ in the single-band system at $T=T_{\rm c}$, $1.2T_{\rm c}$, and $1.4T_{\rm c}$ obtained from the exact analytical continuation from Ref.~\cite{Palestini} as well as the Pad\'{e} approximants (thin curves).
The parameters are set at $(k_{\rm F}a)^{-1}=0$. 
$N_{\rm s,0}=m\sqrt{2mE_{\rm F,s}}/(2\pi^2)$ is the DOS at the Fermi level for a non-interacting Fermi gas at $T=0$.
}
\label{figs1}
\end{figure}
In this Supplemental Material, we show the validity of the Pad\'{e} approximants,
which assume that $\Sigma_i(\bm{p},z)$ with the complex frequency argument $z$ for given $\bm{p}$ is in the form
\begin{eqnarray}
\Sigma_i(\bm{p},z)=\frac{\alpha_1+\alpha_2z+\cdots+\alpha_{j}z^{j-1}}{\beta_1+\beta_2z+\cdots+\beta_{j}z^{j-1}+z^j}.
\end{eqnarray}
The parameters $\{\alpha_k,\beta_k\}$ ($k=1,\cdots,j$) are determined by the $2j$ numerical values of $\Sigma_{i}(\bm{p},i\omega_\ell)$ along the imaginary axis.
In this work, we use $200$ ($=2j$) data.
\par
In the $T$-matrix approach, one can analytically perform the analytic continuation~\cite{Palestini}.
The imaginary part of the retarded self-energy in this approximation can be written as
\begin{eqnarray}
{\rm Im\Sigma}_i(\bm{k},\omega)=-\sum_{\bm{q}}{\rm Im}\Gamma_{ii}(\bm{q},\omega+\xi_{\bm{q}-\bm{k},i})\left[b(\omega+\xi_{\bm{q}-\bm{k},i})+f(\xi_{\bm{q}-\bm{k},i})\right],
\end{eqnarray} 
where $b(x)=[e^{x/T}-1]^{-1}$ and $f(x)=[e^{x/T}+1]^{-1}$ are Bose and Fermi distribution functions, respectively.
The real part of the self-energy can be obtained via the Kramers-Kronig relation
\begin{eqnarray}
{\rm Re\Sigma}_i(\bm{k},\omega)=\frac{1}{\pi}\mathcal{P}\int_{-\infty}^{\infty}d\omega'\frac{{\rm Im}\Sigma_i(\bm{k},\omega)}{\omega'-\omega},
\end{eqnarray}
where $\mathcal{P}$ is the Cauchy principal value.
To see how the Pad\'{e} approximants work in the analytic continuation procedure,
we compare the DOS with the exact analytic continuation in Ref.~\cite{Palestini} and that with the Pad\'{e} approximants in the $T$-matrix approach.
For simplicity, we consider the single-band system ($i={\rm s}$).
Here we define the non-interacting DOS at the Fermi level $N_{0,{\rm s}}=\frac{m\sqrt{2mE_{\rm F,s}}}{2\pi^2}$ where $E_{\rm F,s}$ is the Fermi energy at $T=0$ in the single-band system. 
Figure~\ref{figs1} shows the DOS at unitarity in the single-band system at $T=T_{\rm c}$, $1.2T_{\rm c}$, and $1.4T_{\rm c}$.
The results with the Pad\'{e} approximants represented by the thin curves show an excellent agreement with those with exact analytic continuation done in Ref.~\cite{Palestini} even near $T=T_{\rm c}$.
\par
\begin{figure}[t]
\begin{center}
\includegraphics[width=7cm]{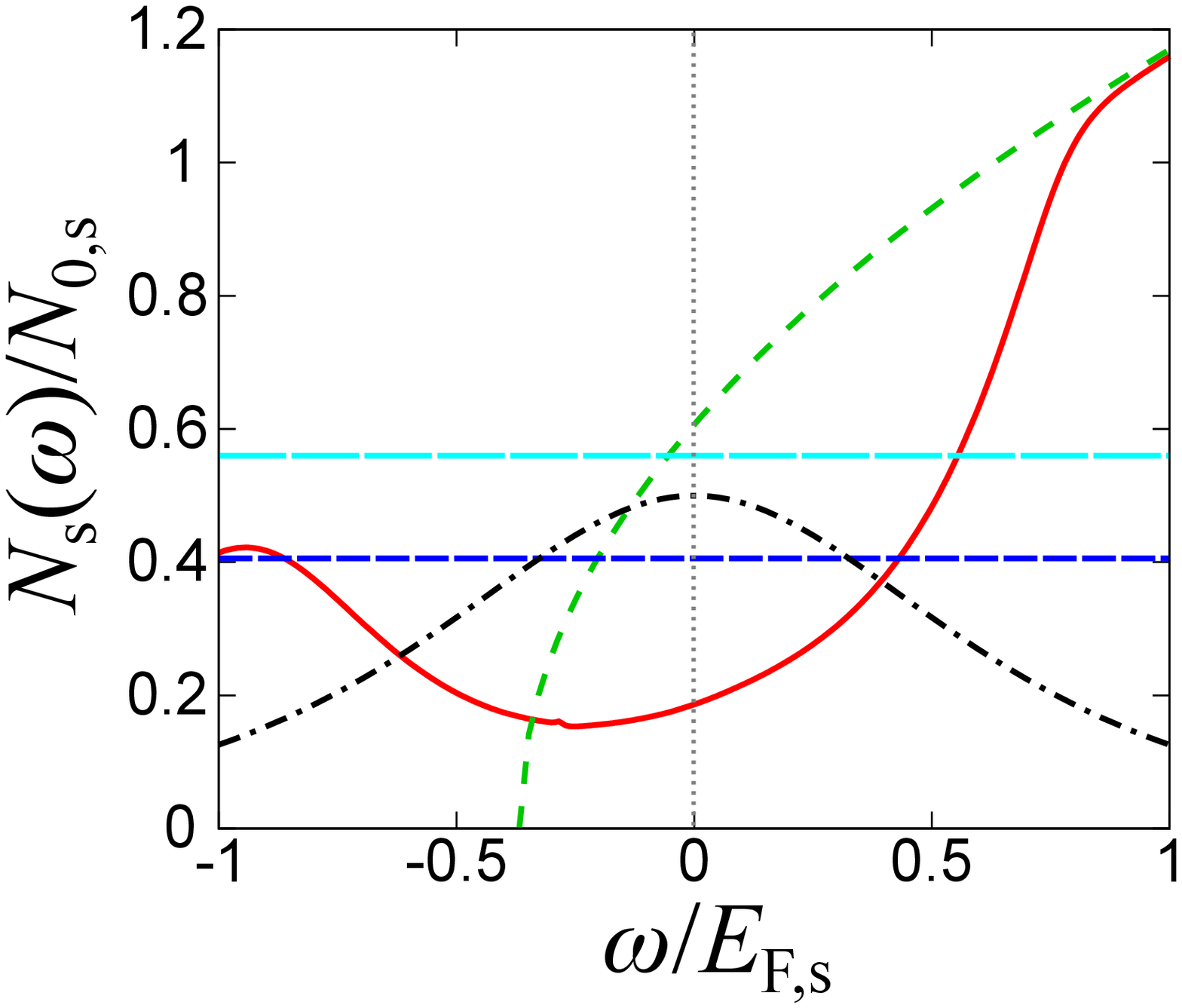}
\end{center}
\caption{Comparison of the DOS $N_{\rm s}(\omega)$ obtained from the Pad\'{e} approximants  (solid curve) and $-G_{\rm s}(\bm{r}=\bm{0},\tau=\beta/2)\beta/\pi=0.405N_{\rm 0,s}$ (dashed line) in the single-band model at $T=T_{\rm c}=0.243T_{\rm F,s}$ in the unitarity limit.
We also plot the square-root type DOS in a non-interacting counterpart and $-G_{\rm s}^0(\bm{r}=\bm{0},\tau=\beta/2)\beta/\pi=0.560N_{\rm 0,s}$ (long-dashed line).
The dash-dotted curve shows the weight factor $1/\left[2\cosh(\beta\omega/2)\right]$ in Eq.~(\ref{eq5}).
}
\label{figs2}
\end{figure}
For comparison, we calculate the single-particle Green's function $G_i(\bm{r},\tau)$ with the spatial position $\bm{r}$ and the imaginary time $\tau$, which is given by
\begin{eqnarray}
G_i(\bm{r},\tau)=T\sum_{\bm{k},i\omega_l}G_i(\bm{k},i\omega_l)e^{i(\bm{k}\cdot\bm{r}-\omega_l\tau)}.
\end{eqnarray}
At sufficiently low temperature, it is related to $N_i(\omega=0)$ as~\cite{Trivedi1995}
\begin{eqnarray}
\label{eq5}
G_i(\bm{r}=0,\tau=\beta/2)&=&-\frac{1}{2}\int_{-\infty}^{\infty}d\omega\frac{N_i(\omega)}{\cosh(\beta\omega/2)}\cr
&=&-\frac{1}{2}N_i(0)\int_{-\infty}^{\infty}\frac{d\omega}{\cosh(\beta\omega/2)}\left[1+\frac{\omega}{N_i(0)}\left.\frac{dN_i(\omega)}{d\omega}\right|_{\omega=0}+\cdot\cdot\cdot\right]\cr
&\simeq& -\frac{\pi}{\beta}N_i(\omega=0),
\end{eqnarray}
where $\beta=1/T$ is the inverse temperature. 
The correction originating from the leading-order term is proportional to $T^2$, which is neglected for simplicity.
We evaluate $G_i(\bm{r}=\bm{0},\tau=\beta/2)$ as
\begin{eqnarray}
\label{eq6}
G_i(\bm{r}=\bm{0},\tau)&=&\sum_{\bm{k}}e^{-\xi_{\bm{k},i}\tau}\left[f(\xi_{\bm{k},i})-1\right]\cr
&&+T\sum_{\bm{k},i\omega_l}\left[G_i(\bm{k},i\omega_l)-G_i^0(\bm{k},i\omega_l)\right]e^{-i\omega_l\tau},
\end{eqnarray}
where the Matsubara frequency sum is evaluated numerically (see Sec.~\ref{secs2}).
\par
First, we consider the single-band case.
Figure~\ref{figs2} shows the comparison between $N_{\rm s}(\omega)$ and $G_{\rm s}(\bm{r}=\bm{0},\tau=\beta/2)\beta/\pi$ where $G_{\rm s}$ is the single-particle Green's function in the single-band Fermi gas. 
In a non-interacting case with same $\mu$ and $T$, we obtain $G_{\rm s}^0(\bm{r}=\bm{0},\tau=\beta/2)\beta/\pi=0.560N_{\rm 0,s}$, which is close to $N_{\rm 0,s}(\omega=0)=m\sqrt{2m\mu_{\rm s}}/2\pi^2\simeq 0.606N_{\rm 0,s}$ where $\mu_{\rm s}$ is the single-band chemical potential.
The difference between them originates from the leading-order correction in Eq.~(\ref{eq5}).
In the strongly interacting case,
we obtain $G_{\rm s}(\bm{r}=\bm{0},\tau=\beta/2)\beta/\pi\simeq0.405N_{\rm 0,s}$.
Although it is smaller than the non-interacting counterpart, it is larger than the result with the analytic continuation with the Pad\'{e} approximants given by $N_{\rm s}(\omega=0)=0.186N_{\rm 0,s}$.
This is also expected to be the leading-order corrections in Eq.~(\ref{eq5}), which involve not only $N_{\rm s}(\omega=0)$ but also $N_{\rm s}(\omega\neq 0)$ multiplied by the weight factor $1/[2\cosh(\beta\omega/2)]$ shown in Fig.~\ref{figs2}.
To see this, we evaluate the same quantity using $N_{\rm s}(\omega)$ obtained from the analytic continuation with the Pad\'{e} approximants, resulting in $G_{\rm s}(\bm{r}=\bm{0},\tau=\beta/2)\beta/\pi\simeq0.408N_{\rm 0,s}$ 
Indeed, it is close to that obtained from Eq.~(\ref{eq6}) with the Matsubara Green's function $G_{\rm s}(\bm{k},i\omega_l)$.
\begin{figure}[t]
\begin{center}
\includegraphics[width=7cm]{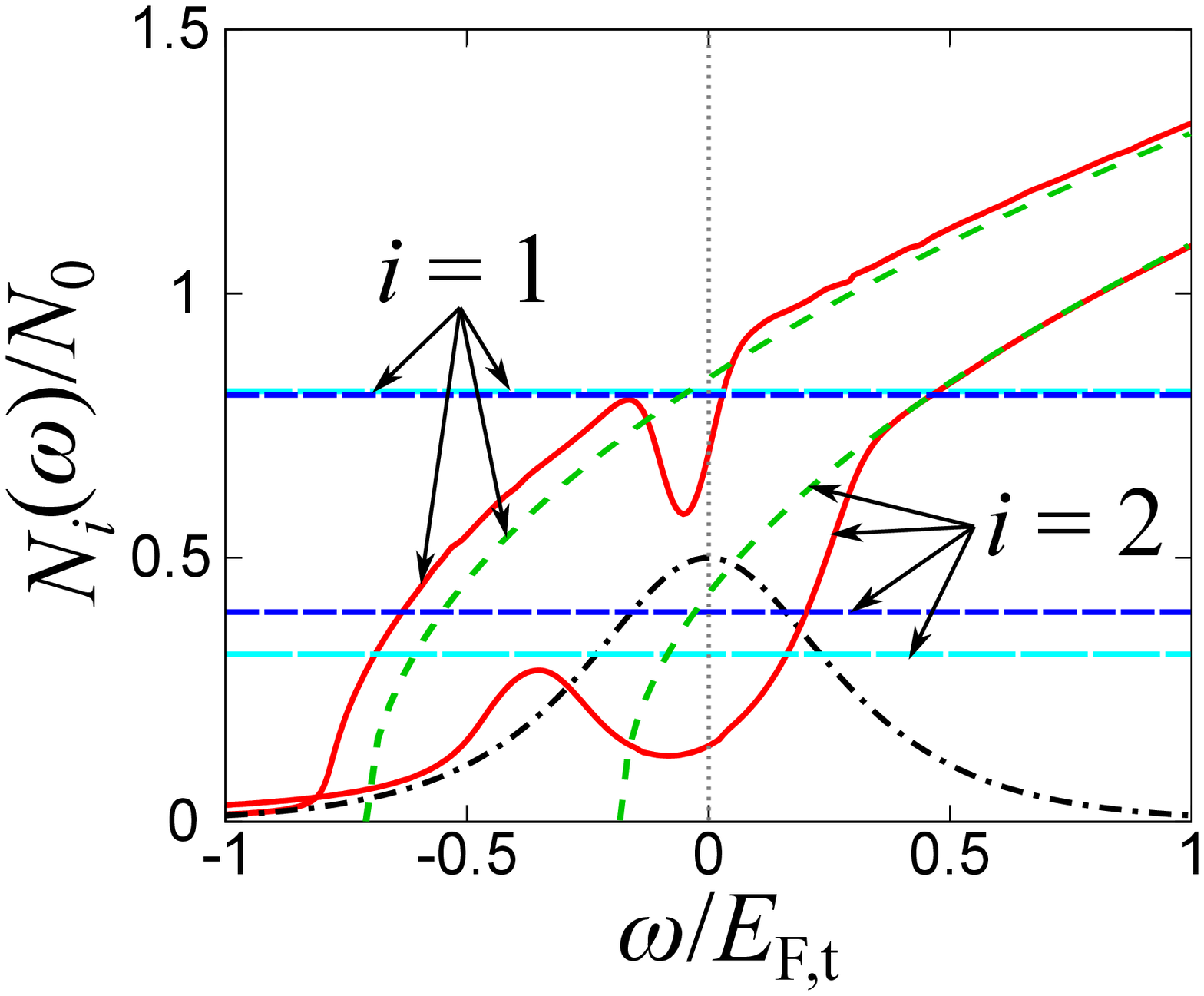}
\end{center}
\caption{Same plots with Fig.~\ref{figs2} in a two-band Fermi gas. The parameters are $T=T_{\rm c}=0.113T_{\rm F,t}$, $(k_{\rm F,1}a_{11})^{-1}=-2$, $(k_{\rm F,2}a_{22})^{-1}=-0.6$, and $\lambda_{12}=2$. The weight factor $1/[2\cosh(\beta\omega/2)]$ (dash-dotted curve) in Eq.~(\ref{eq5}) is also plotted.
}
\label{figs3}
\end{figure}
\par
Figure~\ref{figs3} shows the comparison between $N_{i}(\omega)$ obtained by the analytic continuation with the Pad\'{e} approximants and $-G_{i}(\bm{r}=\bm{0},\tau=\beta/2)\beta/\pi$ in a strongly interacting two-band Fermi gas with $(k_{\rm F,1}a_{11})^{-1}=-2$, $(k_{\rm F,2}a_{22})^{-1}=-0.6$, and $\lambda_{12}=2$ at $T=T_{\rm c}$.
In the weakly-interacting deep band ($i=1$), we obtain $-G_{1}(\bm{r}=\bm{0},\tau=\beta/2)\beta/\pi\simeq0.815N_{0}$ which is close to the non-interacting counterpart given by $0.808N_0$ due to the cancellation of two contributions, that is, the pseudogap suppression and the band-renormalization enhancement of the DOS.
In the strongly-interacting shallow band,
we obtain $-G_{2}(\bm{r}=\bm{0},\tau=\beta/2)\beta/\pi\simeq0.317N_{0}$ which is smaller than the non-interacting counterpart given by $0.397N_0$.
However, it is larger than the results of Pad\'{e} approximants given by $N_2(\omega=0)=0.144N_0$ due to the contribution from $N_2(\omega\neq 0)$.

\section{Matsubara frequency sum}
\label{secs2}
\begin{figure}[t]
\begin{center}
\includegraphics[width=7cm]{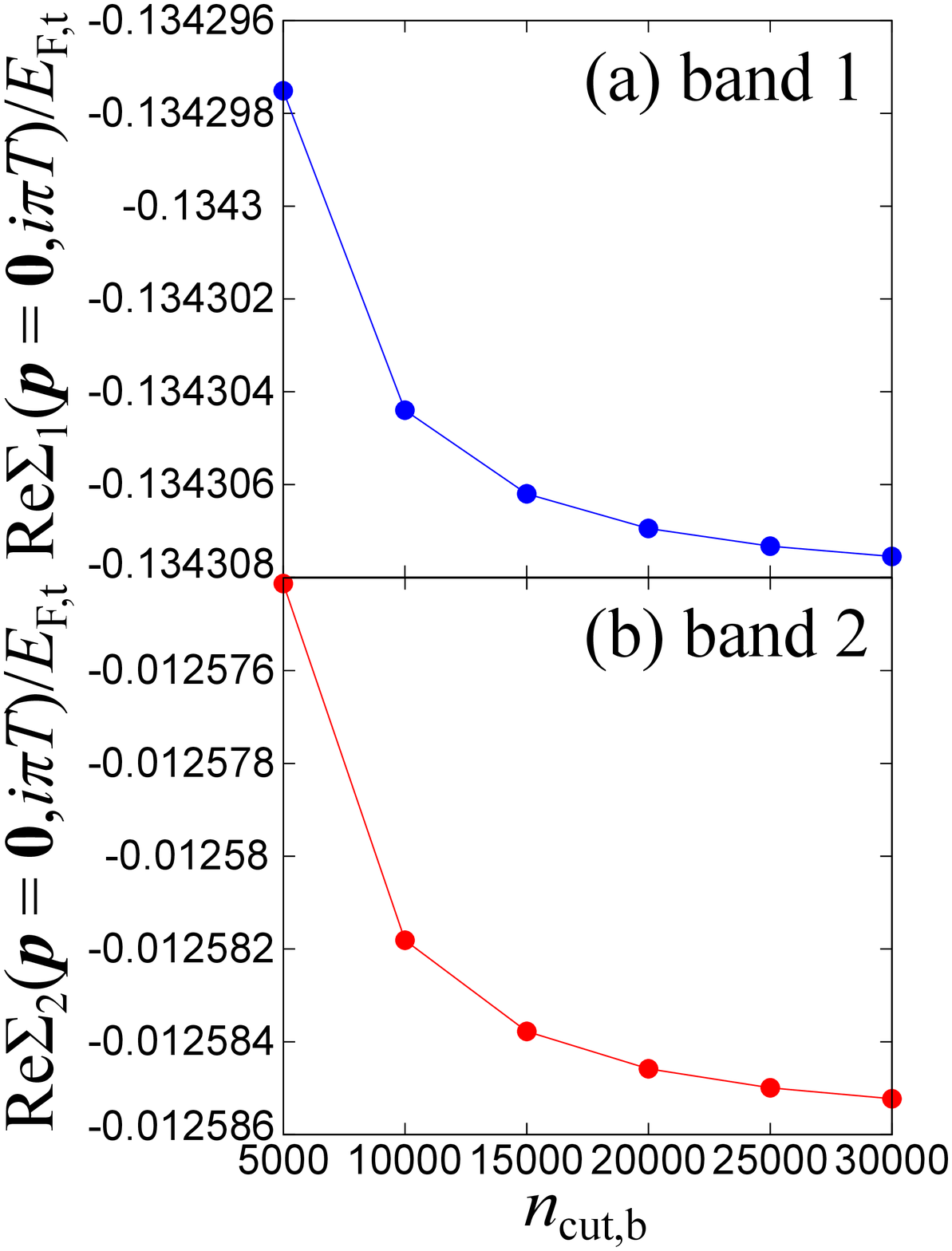}
\end{center}
\caption{The real part of the self-energies ${\rm Re}\Sigma_{i}(\bm{p}=\bm{0},i\omega_l=i\pi T)$ in (a) the deep band ($i=1$) and the shallow band ($i=2$) as a function of the bosonic Matsubara frequency cutoff $n_{\rm cut,b}$ in a two-band Fermi gas at $T=T_{\rm c}$ with $(k_{\rm F,1}a_{11})^{-1}=-2$, $(k_{\rm F,2}a_{22})^{-1}=0$, and $\lambda_{12}=1$.
}
\label{figs4}
\end{figure}
We evaluate numerically the Matsubara frequency sum in the self-energy $\Sigma_i(\bm{p},i\omega_l)$ as
\begin{eqnarray}
\label{eq:sig}
\Sigma_{i}(\bm{p},i\omega_l)&=&U_{ii}n_{i}^0+T\sum_{\bm{p}}\sum_{\ell}^{|\ell|\leq n_{\rm cut,b}}\left[\Gamma_{ii}(\bm{q},i\nu_\ell)-U_{ii}\right]G_i^0(\bm{q}-\bm{p},i\nu_\ell-i\omega_l),
\end{eqnarray}
where $n_i^0$ is the number density for a non-interacting gas and we introduce the cutoff number $n_{\rm cut,b}$.
We take $n_{\rm cut,b}=1000\sim50000$, depending on the coupling parameters as well as the temperature.
In addition, we add the contribution beyond $n_{\rm cut,b}$ by approximately transforming the summation into continuous integration~\cite{Pieri2004}.
In Fig.~\ref{figs4}, we show the dependence by $n_{\rm cut,b}$ of the typical self-energy $\Sigma_{i}(\bm{p}=\bm{0},i\omega_l=i\pi T)$ at $T=T_{\rm c}$ with $(k_{\rm F,1}a_{11})^{-1}=-2$, $(k_{\rm F,2}a_{22})^{-1}=0$, and $\lambda_{12}=1$.
We find sufficient convergences of them within the relative errors of $0.01\%$ in both bands.

\par
We note that the Matsubara frequency sum in $\Pi_{\ell\ell}(\bm{q},i\nu_l)$ can analytically be performed as
\begin{eqnarray}
\label{eq:pi}
\Pi_{\ell\ell}(\bm{q},i\nu_l)=\sum_{\bm{p}}\frac{1-f(\xi_{\bm{p}+\bm{q},\ell})-f(\xi_{\bm{p},\ell})}{i\nu_l-\xi_{\bm{p}+\bm{q},\ell}-\xi_{\bm{p},\ell}}.
\end{eqnarray}

In the case of the number density $n_i$, we decompose the equation with the non-interacting density $n_i^0$, the NSR correction $\delta n_i^{\rm NSR}$~\cite{TajimaCM2020}, and the remaining part $\delta n_i$ as
\begin{eqnarray}
n_i&=&2\sum_{\bm{k}}f(\xi_{\bm{k},i})+2T\sum_{\bm{k},i\omega_n}\left\{G_i^0(\bm{k},i\omega_n)\right\}^2\Sigma_i(\bm{k},i\omega_n)\cr
&&+2T\sum_{\bm{k},i\omega_n}\left[G_i(\bm{k},i\omega_n)-G_i^0(\bm{k},i\omega_n)-\left\{G_i^0(\bm{k},i\omega_n)\right\}^2\Sigma_i(\bm{k},i\omega_n)\right]\cr
&\equiv&n_i^0+\delta n_{\rm NSR}+\delta n_i.
\end{eqnarray}
Using the same technique in Eq. (\ref{eq:pi}), we can analytically perform the fermionic Matsubara summation in $\delta n_{i}^{\rm NSR}$ as~\cite{TajimaCM2020}
\begin{eqnarray*}
\label{eq:nsr}
\delta n_i^{\rm NSR}=-T\sum_{\bm{q},i\nu_l}\frac{U_{ii}[1+U_{\bar{i}\bar{i}}\Pi_{\bar{i}\bar{i}}(\bm{q},i\nu_l)]-U_{12}U_{21}\Pi_{\bar{i}\bar{i}}(\bm{q},i\nu_l)}{[1+U_{11}\Pi_{11}(\bm{q},i\nu_l)][1+U_{22}\Pi_{22}(\bm{q},i\nu_l)]-U_{12}U_{22}\Pi_{11}(\bm{q},i\nu_l)\Pi_{22}(\bm{q},i\nu_l)}\frac{\partial \Pi_{ii}(\bm{q},i\nu_l)}{\partial \mu},
\end{eqnarray*}
where $\bar{i}$ denotes the opposite band index of $i$ (e.g. $\bar{i}=1$ when $i=2$). 
We note that the bosonic Matsubara frequency sum in Eq.~(\ref{eq:nsr}) is numerically evaluated with the same technique used for the self-energy calculation in Eq.~(\ref{eq:sig}).
\par
When we  perform the fermionic Matsubara sum in $\delta n_i$, we introduce the cutoff number $n_{\rm cut,f}$ as
\begin{eqnarray}
\delta n_i=2T\sum_{\bm{p}}\sum_{n}^{|n|\leq n_{\rm cut,f}}\left[G_i(\bm{k},i\omega_n)-G_i^0(\bm{k},i\omega_n)-\left\{G_i^0(\bm{k},i\omega_n)\right\}^2\Sigma_i(\bm{k},i\omega_n)\right],
\end{eqnarray} 
in which the convergence with respect to $n_{\rm cut,f}$ is faster compared to the summation of $G_i(\bm{k},i\omega_n)$ without the decomposition.
\begin{figure}[t]
\begin{center}
\includegraphics[width=7cm]{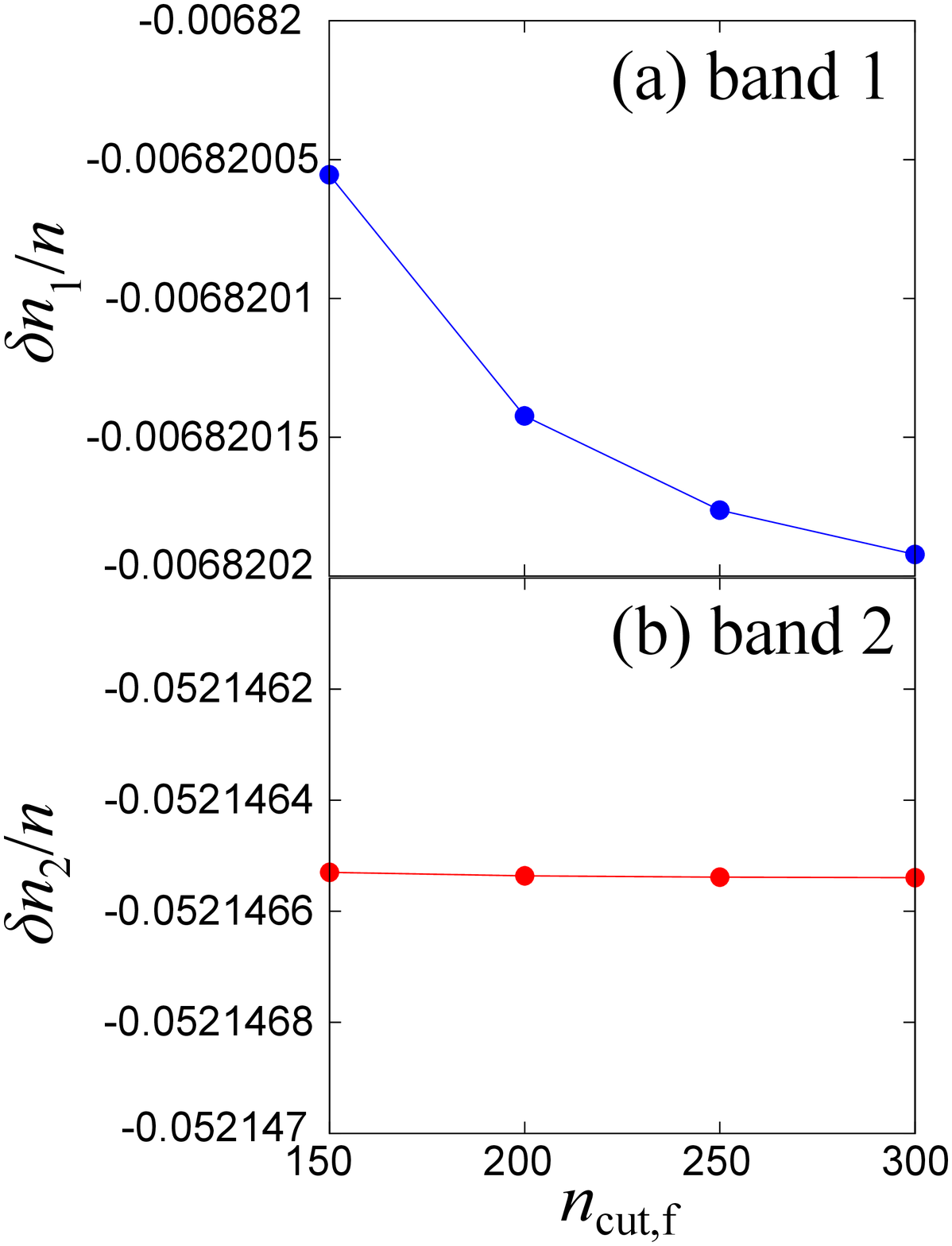}
\end{center}
\caption{ 
The correction beyond the NSR approach to the number densities (a)$\delta n_{1}$ and (b)$\delta n_{2}$ as a function of the fermionic Matsubara frequency cutoff $n_{\rm cut,b}$ in a two-band Fermi gas at $T=T_{\rm c}$ with $(k_{\rm F,1}a_{11})^{-1}=-2$, $(k_{\rm F,2}a_{22})^{-1}=0$, and $\lambda_{12}=1$.
}
\label{figs5}
\end{figure}
Figure~\ref{figs5} shows the $n_{\rm cut,f}$ dependence of $\delta n_i$
in a two-band Fermi gas at $T=T_{\rm c}$ with $(k_{\rm F,1}a_{11})^{-1}=-2$, $(k_{\rm F,2}a_{22})^{-1}=0$, and $\lambda_{12}=1$.
We find sufficient convergences for $n_{\rm cut,f}$ at each coupling parameter and temperature.
We use $n_{\rm cut,f}=200\sim 300$, checking their convergences within the relative errors of $0.01\%$.

\end{document}